\documentclass[
aps,%
11pt,%
final,%
notitlepage,%
oneside,%
twocolumn,%
nobibnotes,%
nofootinbib,%
superscriptaddress,%
noshowpacs,%
centertags]%
{revtex4}

\begin{document}
\title{Analysis of a Sample of RC Catalog Objects in the Region Overlapping
with the Areas Covered by FIRST and SDSS Surveys. II: Optical Identification
with the SDSS Survey and USNO-B1 and 2MASS Catalogs}
\author{\firstname{O.~P.}~\surname{Zhelenkova}}
\affiliation{\saoname}
\author{\firstname{A.~I.}~\surname{Kopylov}}
\affiliation{\saoname}

\begin{abstract}
We report the results of optical identification of a sample of RC catalog
radio sources with the FIRST and SDSS surveys. For 320 sources identified
with  NVSS and FIRST objects we perform optical identification with the SDSS
survey. When selecting optical candidates we make maximum use of the
information about the structure of radio sources as provided by the FIRST
survey images. We find optical candidates for about $80$\% of all radio
sources.
\end{abstract}
\maketitle

\section{INTRODUCTION}

Powerful radio galaxies can be observed almost at any cosmological distance,
however, their host galaxies are often optically faint. Much observational
time is needed to identify a radio source and find the redshift of its host
galaxy. The history of the identification of the 3CR  catalog of bright radio
sources ($\nu$=187\,MHz, $S_{lim}$=5\,Jy)
\cite{Bennet,Laing} serves to illustrate this
point. The limiting magnitude of the Palomar Observatory Sky Survey (POSS)
of about $20.5^m$ for E plates, proved to be sufficient to allow the
identification of 65\% of all radio sources of the catalog, whereas it took
almost three decades to identify the remaining  35\%
\citep{Kristian,Pooley,Rawlings,deKoff,McCarthy,Martel}.
The number of optically identified radio sources rapidly decreases at lower
fluxes. For example, the fraction of POSS identifications for B2 survey
($\nu$=408\,MHz, $S_{lim}=250\,$mJy) is equal to 38\% \cite{Grueff},
whereas the corresponding fraction for deeper the First ($\nu$= 610\,MHz,
$S_{lim}\sim20$mJy) and the Second ($\nu$=1415\,MHz, $S_{lim}\sim7$mJy)
Westerbork surveys is even lower, of about 20\% \cite{Katgert}.

The field of view of a large telescope does not exceed $10\arcmin-20\arcmin$,
making the task of finding enough reference stars for the astrometric
calibration problematic at the times before the release of the
APM~\cite{Irwin}, USNO \cite{Monet}, and GSC \cite{Morrison} catalogs.
For this reason, secondary astrometric standards had to be used in the fields
studied.

Extended radio sources with complex structure are identified using composite
images made up of optical and radio frames. A reliable choice of the optical
candidate requires accurate coordinate calibration (to within $1\arcsec$),
high-resolution radio images (e.g., like those of the FIRST survey with an
angular resolution of 5.4$\arcsec$), and deep direct optical images (with an
R-band limiting magnitude of $23^m$--$24.5^m$) with the astrometric
calibration accuracy at least as good as that of the radio image.
If even after this procedure there remain several optical candidates,
additional information, e.g., broad-band photometry, can be used to pick up
the  right one. However, only spectroscopic observations allow the
host galaxy of the radio source to be reliably identified.

In early 1980s a deep survey was carried out with the \mbox{RATAN-600} within
a $20\arcmin$-wide sky strip centered at the declination of SS433
($\delta_{1950.0} = +4\degr54\arcmin$). The survey angular resolution at
3.94\,GHz ($\lambda$=7.6\,cm) was equal to $\Delta\alpha\sim1\arcmin$. The
RC catalog compiled based on the observation data of this survey has a
coordinate accuracy of $5\arcsec \times 45\arcsec$ in right ascension and
declination, respectively \cite{Berlin1,Berlin2}. The faintest sources
detected have flux densities of about $4$\,mJy. The completeness of the
catalog is equal to 0.8 for sources with flux densities
$S_{3.94\,GHz}$>7.5\,mJy within the $\pm5\arcmin$ strip about the central
declination of the survey. The RC catalog contains a total of  1165
sources~\mbox{\cite{Parijskij,Par}}.

A pilot program of optical identification of objects of the RC catalog was
performed using the photographic data obtained in 1984--1989 with the 6-m
telescope of the Special Astrophysical Observatory of the Russian Academy of
Sciences~\cite{Vitkovskij}. It was followed by the identification of a
266-object sample with the refined coordinates based on the  TXS
catalog~\cite{Douglas} via inspection of enlarged POSS \mbox{prints
\cite{Soboleva,Fletcher}.} Optical candidates have been found
for 72 ($\sim27\%$) of these objects.

The next stage of the work involved the identification of a sample of
steep-spectrum radio sources of the RC catalog on deep CCD images
(\mbox{$m_{R}^{lim}\sim24^m$}) taken with the 6-m telescope
within the framework of the ``Big Trio'' program of the search for distant
radio galaxies \cite{Kopylov,Parijskij1,Verkhodanov}.
The astrometric calibration of small regions of CCD frames
\mbox{(about $3\arcmin\times3\arcmin$)} was based
on secondary standards whose coordinates were determined using POSS-I
images~\cite{Lasker} and  APM and GSC catalogs. Identification was performed
by means of overlapping high-resolution VLA radio images with optical images.

The release of the deep optical SDSS sky survey~\cite{Abazajian} and the
FIRST radio survey \cite{Becker} made it possible to continue the
identification of RC objects and refine the results of the identification of
radio sources performed by Soboleva et al.~\cite{Soboleva}
and Fletcher et al.~\cite{Fletcher}. About $50\%$ of FIRST radio sources are
estimated to be identified down to the limiting magnitude of the SDSS survey
($m_{r}=22.6^m$), and we therefore believed that at least the same fraction
of RC objects can be identified with SDSS.

\section{OPTICAL IDENTIFICATION}

The RC catalog overlaps with the SDSS and FIRST surveys in the
right-ascension interval from $8^h 11^m$ to $16^h 25^m$, which includes a
total of  432 objects. A comparison of the images of the NVSS survey
\cite{Condon} with angular resolution of  $45\arcsec$ with the FIRST survey
images with an angular resolution of  $5.4\arcsec$ shows that the latter
should be used for optical identification, especially in cases where the
corresponding  NVSS object is not a point source
(angular size exceeds $23\arcsec$) and breaks into independent sources in
the FIRST images.

In complex cases  (multicomponent sources, groups of independent radio
sources, or several optical candidates) detailed information about the
structure of the source is required for optical identification. To establish
the structure of the radio sources according to the FIRST survey data in
more detail, we not only drew the contours using the Aladin
application~\cite{Bonnarel}, but also employed a software
tool that draws flux density isophotes in the areas studied without the loss
of angular resolution~\cite{Service}.

The structure of the radio source correlates with the position of the host
galaxy, and therefore not only on the coordinate coincidence, but also on
the morphological type of the source should be taken into account when
choosing the optical candidate. We based our morphological classification of
radio sources on a slightly modified variant of the scheme proposed by
Lawrence et al.~\cite{Lawrence}. Our classification includes
the following types:
\begin{itemize}
\item
point radio sources (core or C). The positions of their host galaxies most
likely coincide with the flux density peak;
\item
sources with one-sided jets  (core-jet or CJ). For such sources, the position
of the optical candidate coincides with that of the bright compact component;
\item
sources with a bright core and components (core-lobe or CL). In these sources
intensity of radio emission decreases from the center toward the periphery.
The optical candidate coincides with the peak of the intensity distribution;
\item
double sources (double or D) and double sources with double components
(double-double or DD). The optical candidate in double radio sources is
located between the components or coincides with the minimum of the intensity
distribution between the merging components;
\item
double sources with a core (double-core or DC). A double source with a faint
core located between the components. The optical candidate coincides with
the core;
\item
triple sources (T). The radio source consists of three components, where the
flux of the central part is comparable to those of the other two components.
The optical candidate coincides with the central component;
\item
multiple sources (M). These are multicomponent sources with a structure that
is difficult to put into any of the above categories. In this case the
position of the optical candidate is not easy to determine and additional
photometric or spectroscopic information is needed for a choice of the
optical objects located within the error box of the coordinate calibration.
\end{itemize}

We  classified 320 sources in accordance with this scheme. We then determined
the position of the optical candidate based on the morphological type of the
source. In a number of cases its position with respect to the isophotes of
the radio image made it possible to refine the type of the radio
source or identify the components with individual radio sources. For example,
if each of the components of the presumed double source had the size of about
$2\arcsec$ and coincided with a certain optical object, we regarded it as
two point sources, because accidental coincidence is highly unlikely  due
to the low surface density of the radio sources.

We took the following coordinates for a intended optical candidate (hereafter
referred to as the center of the radio source):
\begin{itemize}
\item
F---the coordinates of the point source or of the source with a well-defined
core as listed in the FIRST catalog (213 sources);
\item
N---if the source has two components with different flux densities in the
FIRST survey, which constitute a single object in the NVSS, we adopt the
NVSS coordinates, which correspond to the location of the center of mass
the radio source, as the position of the optical object (32 sources);
\item
Fm---the coordinates of the center of the double source as measured by the
contour map of the image in the FIRST survey (75 sources) in the following
cases:
     \begin{itemize}
     \item[(1)]
     if the image contains a faint core with coordinates that are not listed
     in the FIRST catalog,
     \item[(2)]
     in the absence of a core the presumed position is determined by the
     contours of the FIRST radio image.
\end{itemize}
\end{itemize}

We consider the optical object to be the most likely candidate for
identification (``+'') if it was located (according to the SDSS catalog)
within $3\sigma$ of the center of the radio source, where $\sigma$ is the
coordinate error. We consider identifications to be doubtful (``?'')  in
the following cases:
\begin{itemize}
\item
if the object is a point source and the possible optical counterpart, albeit
close, lies beyond $3\sigma$ from the center of the radio source;
\item
if there are two optical objects in the vicinity of the expected location of
the optical candidate;
\item
if the source is double, the position of its core is uncertain, and the
optical object is offset from the line connecting the brightness maxima of
the lobes;
\item
if it is difficult to make any definitive conclusions about the structure of
the radio source based on the FIRST radio map (whether it should be viewed
as a multicomponent radio source or several independent radio sources).
\end{itemize}

The average coordinate error of the  FIRST survey is $0.5\arcsec$, amounting
to 1$\arcsec$  for $S_{1.4 GHz} \sim1$mJy sources. The accuracy  of the radio
coordinates is better than 1$\arcsec$ for NVSS survey sources with flux
densities $S_{1.4 GHz}$>15\,mJy.
We estimate the coordinate accuracy for the objects with the central part of
the radio source identified using method ``Fm'' or ``N''
to be  1$\arcsec$, and that for the objects with the central part of the
radio source identified using method ``F'' to be $0.5\arcsec$. In the sky
area studied the average density of SDSS objects  (for $m_{r}^{lim}=22.6^m$)
is about $0.0020/\square\arcsec$. The accuracy of the coordinate calibration
of SDSS is about $0.1\arcsec$.

We use these values to compute the normalized distance $D$ between the
optical candidate and the center of the radio source:
$$
D=\sqrt{\frac{\Delta\alpha^2}{\sigma_{\alpha}^2}+\frac{\Delta\delta^2}{\sigma_{\delta}^2}}
\mbox{ , where }
\sigma_{\alpha}^2=\sigma_{\alpha_{rad}}^2+\sigma_{\alpha_{opt}}^2
$$
 and
$\sigma_{\delta}^2=\sigma_{\delta_{rad}}^2+\sigma_{\delta_{opt}}^2$.

We use the maximum-likelihood ratio $LR$\footnote{
See \cite{Best} for a detailed description of the procedure employed to
compute the probability, reliability, and completeness of optical
identification for point and double radio sources. We adopted
the above formulae from that paper.
}
as the effective estimate for the reliability of identification. We compute
$LR$ for point sources by the following formula:
$$
LR(D)=\frac{1}{2\lambda}e^{\frac{D^2}{2}(2\lambda-1)}
\mbox{, where\,} \lambda=\pi\sigma_{\alpha}\sigma_{\delta}\rho
$$
and $\rho$ is the number of objects per square arcsecond.

\begin{table*}[p!]
\setcaptionmargin{0mm} \onelinecaptionstrue
\caption{The quantities that characterize the reliability of optical
identification for objects of various types: the mean, root-mean-square
deviation and median of the separation between the optical candidate and the
radio source; the cutoff maximum-likelihood ratio, and the corresponding
cutoff search radius. No data are available for four radio sources (two
multicomponent objects and two objects without  FIRST survey maps)
}
\label{lr}
\begin{center}
\begin{tabular}{ l | c | c | c | c | c | c | c  }
\hline
Type        & $N_{obj}$  & ~$\phi=N_{+}/N_{obj}$~ & ~$\Delta d_{rad-opt}^{mean}$,~ & ~$\sigma_{\Delta d_{rad-opt}}$,~ & ~$\Delta d_{rad-opt}^{med}$,~ & ~$LR_{cutoff}$~ & ~$\Delta d_{rad-opt}^{cutoff}$,~ \\
       &            &                        & ~arcsec                  & ~arcsec                    & ~arcsec                 &                 & ~arcsec                    \\
\hline
~C           & ~124 (39\%)~ & 0.64   & 0.94        & 1.30        & 0.47        & 1.79          & 1.88    \\
~D, DD        & ~108 (34\%)~ & 0.72   & 1.43        & 0.87        & 1.26        & 1.93          & 3.49    \\
~DC, CJ, CL, T~ & ~84 (26\%)~  & 0.76   & 1.44        & 2.16        & 0.75        & 1.41          & 3.58    \\
\hline
\end{tabular}
\end{center}
\end{table*}

The above formula can be used to compute $LR$ if the position of the host
galaxy can be reliably found from the structure of the radio source, which is
the case for the CJ, CL, DC, and T types described above (65\% of all sources
in our sample). We also use this formula to compute the $LR$ values for
double radio sources in the cases where the supposed position of the core can
be reliably determined from the structure of the components. This is true for
almost  2/3 of all double sources in our sample.

An empirical relation between the reliability and completeness of
identification for radio sources \cite{Best} as a function of $LR$ determines
the cutoff value ($LR_{cutoff}$), which separates identification from
eventual chance coincidence with background objects.

The radio sources for which the optical candidates have been found can be
subdivided into the following three groups by the morphological type and
center of the radio source:
\begin{itemize}
\item
point sources (C),
\item
sources where the peak of radio emission coincides with the optical object
(CL, CJ, DC, and T), and
\item
double radio sources (D and DD), where the supposed position of the optical
object was determined based on the shape of the isophotes of the radio image
and not by the coordinates listed in the FIRST catalog.
\end{itemize}
The cutoff levels of the maximum-likelihood ratios, $LR_{cutoff}$, and the
corresponding offsets $d_{rad-opt}$ between the optical and radio coordinates
proved to be more or less the same for all the three groups
(see Table~\ref{lr}), thereby lending a certain support to our proposed
procedure of searching for optical candidates.

\begin{figure*}[tbp]
\setcaptionmargin{5mm}
\onelinecaptionsfalse
\includegraphics[width=15cm]{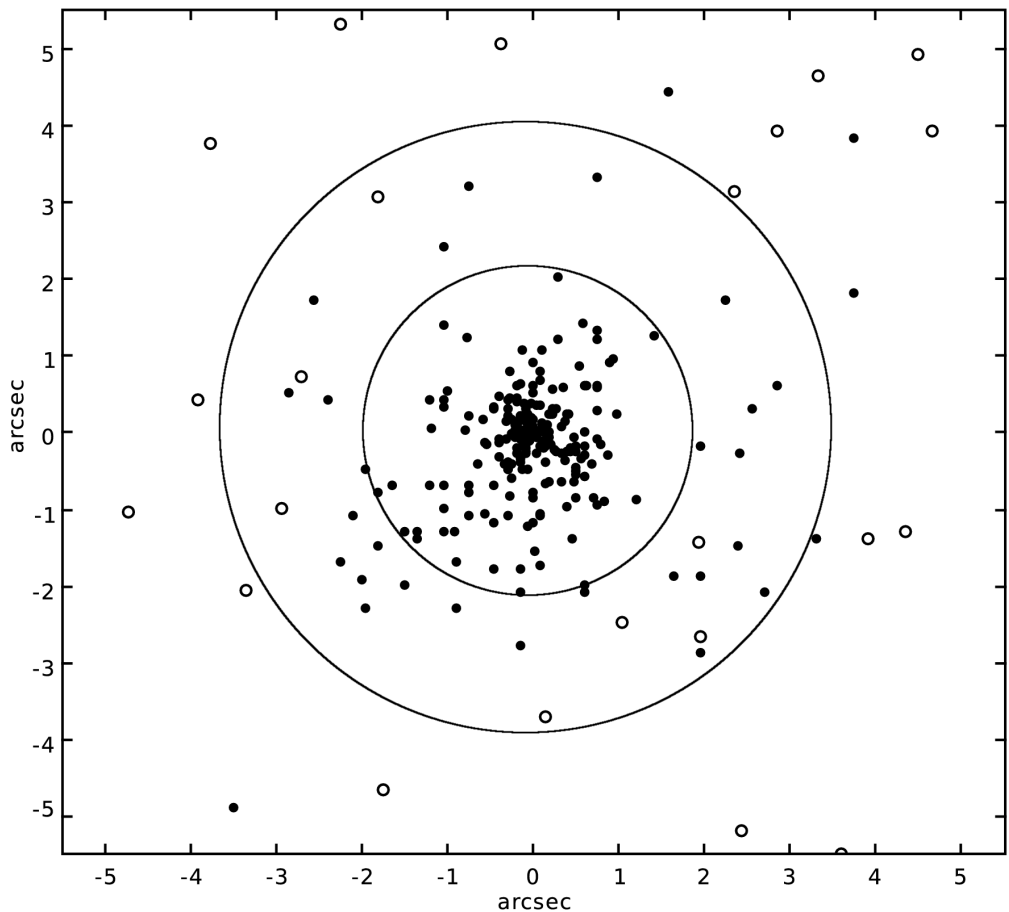}
\captionstyle{normal}
\caption{
Distribution of the offset between the radio and optical coordinates for RC
objects. The black dots and circles correspond to bona fide  (``+'') and
likely  (``?'') optical identifications, respectively. The source
coordinate differences in right ascension and declination are plotted along
the x- and y-axis, respectively. The inner and outer circles are the
boundaries of the coordinate-offset domains corresponding to  $LR_{cutoff}$
for point sources and sources with a core, respectively.
   }\label{optrad}
\end{figure*}

We use the computed $LR$ ratios for the optical candidates and the
$LR_{cutoff}$ values as a testing criterion in the selection of the most
likely identification. Figure~\ref{optrad} shows the distribution of the
offsets between the radio and optical coordinates for the ``+'' and ``?''
optical candidates.

There are gaps in the equatorial area of the SDSS survey. Inside these gaps
we searched for optical candidates of radio sources in the USNO-B1 catalog
and identified one source (RC~J1623+0446) in such a way.

We found another source (RC~J1052+0458) to have an optical candidate in the
SDSS images, but lack the data  in the SDSS database. This object is listed
in the  USNO-B1 ($R2=19.45^{m}$) and 2MASS catalogs \cite{Cutri}
($K=14.99^{m}$).

Below we give a few examples of the identification of radio sources and the
reasoning used to select the optical candidates.

We found two optical candidates for the RC~J1257+0458 source:
a galaxy (SDSS type ``GALAXY'') and a starlike object (``STAR''). By its  u,
g, r, i, and z-band magnitudes the latter looks more like a star rather than
a quasar (quasars, too, are classified as ``STAR''). We chose the galaxy as
the optical candidate.

In the first our paper~\cite{Zhel} dedicated to the identification of the RC
catalog with the VLSS, TXS, NVSS, FIRST, and GB6 radio catalogs we give an
identification for the double radio source RC~J0815+0453.
It is evident from the radio isophotes that  RC~J0815+0453 has a
well-defined core, which exactly coincides with the galaxy and hence the
latter is the true optical identification rather than the fainter object
located precisely on the line connecting the intensity maxima of the radio
source lobes. There is yet another faint FIRST~J081521.3+045339
radio source ($S^{peak}_{1.4\,GHz}$=2.67\,mJy) near  RC~J0815+0453. Both
radio sources are identified with elliptical galaxies with practically the
same photometric redshifts. Thus, according to SDSS data, the
photometric redshift of the host galaxy of the fainter radio source is
\mbox{$Z_{phot1}=0.0933\pm0.0049$} (based on the approximation by pattern
spectra~\cite{Csabai}) and \mbox{$Z_{phot2}=0.0697\pm0.0142$}
(neural-network meth\-od~\cite{Collister}). The corresponding photometric
redshift estimates for the second galaxy are $Z_{phot1}=0.1235\pm0.0052$ and
\mbox{$Z_{phot2}=0.0694\pm0.0149$,} respectively.
It seems fair to suppose---although this requires further studies---that the
two sources form a close, possibly an interacting, pair of radio galaxies.

\begin{figure*}[tbp]
\setcaptionmargin{5mm}
\onelinecaptionsfalse
\includegraphics[width=0.7\textwidth,bb=39 207 506 617,clip]{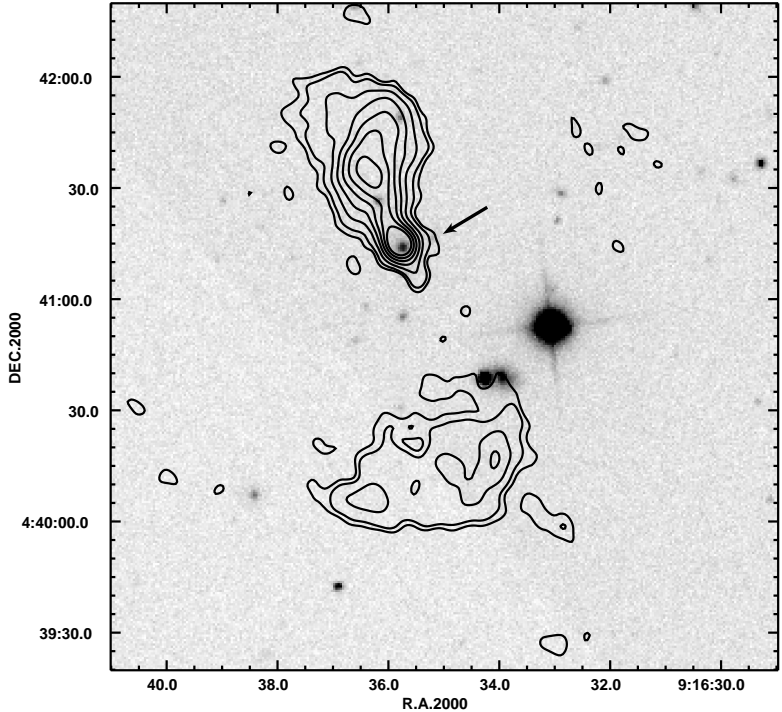}
\captionstyle{normal}
\caption{
A composite optical (SDSS) and radio (FIRST) image of the  RC J0916+0441
radio source. The isophotes drawn preserving the angular resolution show
that both the Northern and Southern sources are components of a single
``core-lobe'' type radio source. The arrow indicates the position of the
host galaxy.
   }\label{RC0916}
\end{figure*}

The identification of  RC~J0916+0441 required a detailed information on its
structure. Thus one might conclude, based on the isophotes of the FIRST
survey image produced by Aladin application, that RC~J0916+0441
should be a blend of two independent radio sources---the Northern source
consisting of two objects of the  FIRST catalog and the Southern source
consisting of four sources of the FIRST catalog.
However, the isophotes drawn with angular resolution preserved (we use the
software service \cite{Service} to this end) allowed us to understand the
structure of the source. It proved to be an object of morphological
type ``core-lobe'' with the core coincident with a galaxy
(see Fig. ~\ref{RC0916}).

\begin{figure*}[tbp]
\setcaptionmargin{5mm}
\onelinecaptionsfalse
\includegraphics[width=0.7\textwidth,bb=40 201 506 622,clip]{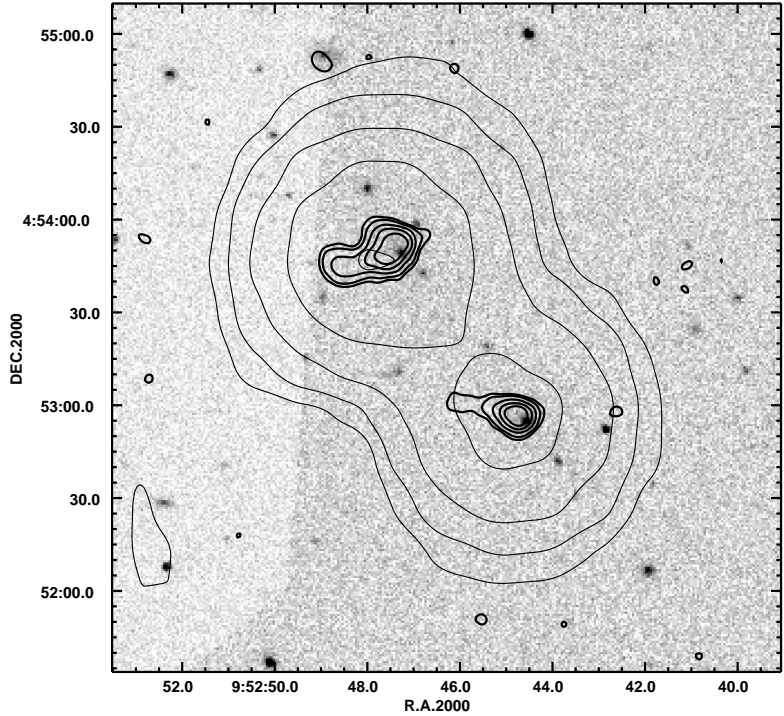}
\captionstyle{normal}
\caption{
The RC~J0952+0453 source is located between the two components of an NVSS
source (indicated by thinner contours).  The higher angular resolution of
the FIRST survey allows RC J0952+0453 to be resolved into two separate radio
sources, because the isophote contours are unconnected, the orientations of
the major axes do not coincide, and each of the components of the radio
source is identified with a separate  optical object.
   }\label{RC0952}
\end{figure*}

We show our next example in Fig.~\ref{RC0952}. Here RC~J0952+0453 is located
between the components of a double NVSS source, closer to the brighter
component. It is evident from the shape of their isophotes as shown on the
FIRST contour map that the source components  are unconnected, their
orientation is not coaxial, and each component can be identified with a
separate optical object.
The RC~J0952+0453 source is thus a blend of two independent radio sources.

In similar cases, where the size of the beam of RATAN-600 does not make it
possible to resolve close radio sources, we assumed the brightest component
to be the main contributor to the RC catalog source and then proceeded to
its optical identification.

\section{RESULTS OF IDENTIFICATION}

\begin{figure*}[tbp]
\setcaptionmargin{5mm}
\onelinecaptionsfalse
\includegraphics[width=15cm]{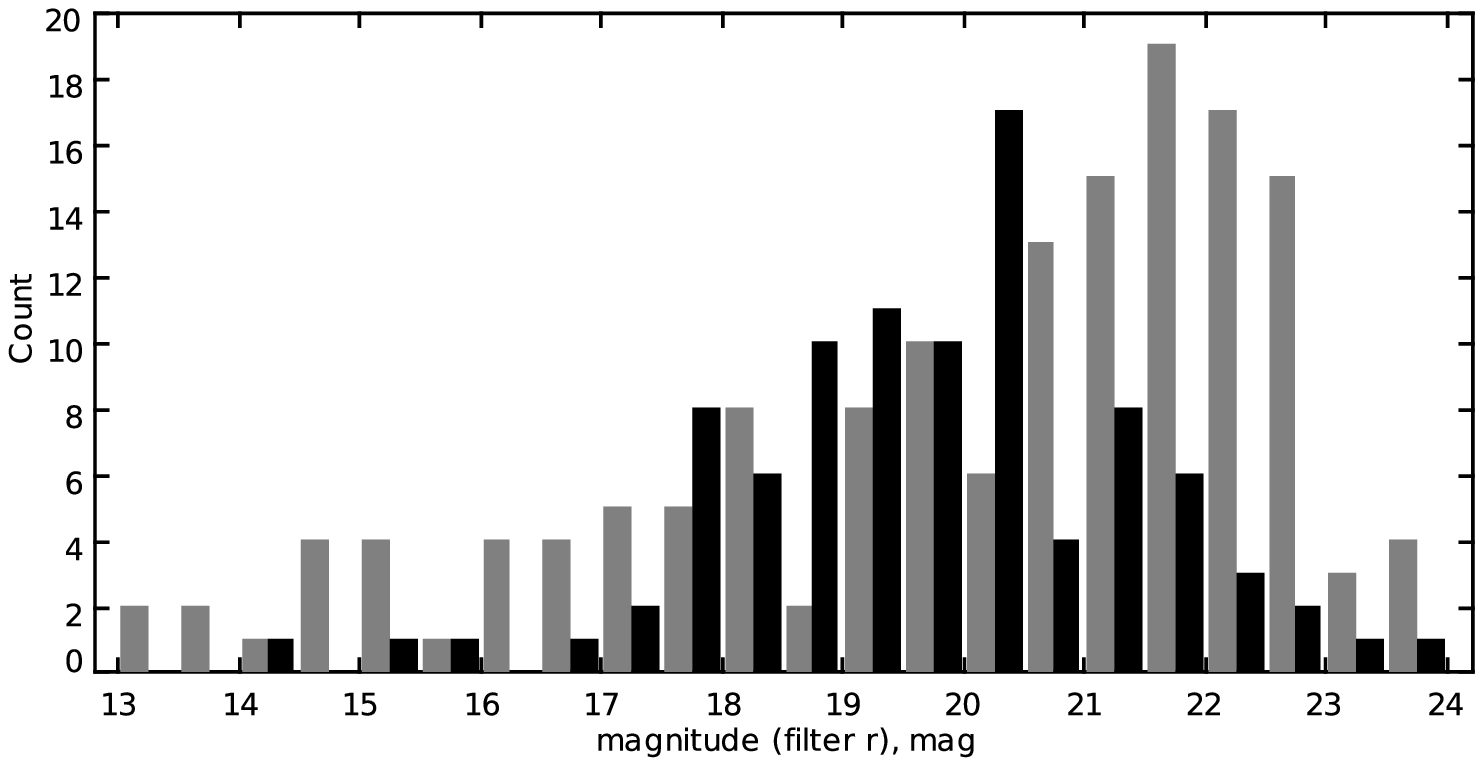}
\captionstyle{normal}
\caption{Distribution of the r-band magnitudes of optical objects identified
with the RC catalog radio sources. The distributions for galaxies and stellar
objects are shown by gray and black color, respectively.
   }\label{igs}
\end{figure*}

In our first paper \cite{Zhel} we identified  320 of a total of 432 sources
of the RC catalog located inside the area of intersection of the SDSS and
FIRST surveys.
A visual inspection of SDSS images superimposed on FIRST contour maps using
refined coordinates allowed us to find bona fide  (``+'') optical candidates
for 227 (71\%) of the radio sources; likely  (``?``)
optical candidates for 25 (8\%) of the radio sources, with no optical
candidates to be found for 68 (21\%) of all radio sources.

The area considered contains  36 objects of the sample of RC sources with
steep spectra (SS). We earlier identified 35 sources using the 6-m telescope
CCD images within the framework of the ``Big Trio'' program. For 10 of these
objects we found no optical candidates in the SDSS survey.

Optical candidates (``+'' and ``?'') were found for 264 (82\%) radio sources
out of 320 objects. These include the SS sample and MG~J1131+0456\footnote{
The RC~J1131+0455 source turned out to be the gravitational lens
MG~J1131+0456, which was earlier identified by Hewitt et al.~\cite{Hewitt},
Tonry et al.~\cite{Tonry}, and Kochanek et al.~\cite{Kochanek} using deeper
images than those provided by SDSS.
}. Note that the summary statistical data below does not include the 11
objects identified using deeper images.

Of all the radio sources identified with SDSS, 89 are stellar objects
(``STAR''), 158 are galaxies (``GALAXY''), and six faint objects are
classified as ``UNKNOWN'' in the SDSS database. Stellar objects are generally
brighter (see Fig. ~\ref{igs}) and bluer than galaxies.

Table~\ref{gsef} lists the results of identification of radio sources for
different types of optical candidates depending on the
$\alpha_{1.4-4.85\,GHz}$ spectral index. We found the sources most
difficult to identify to be those with ultrasteep spectra and faint objects
lacking the spectral-index estimates (about $35$\%). The fraction of
unidentified objects with inverse, flat, and steep spectra is
equal to  16--17\%. The fraction of steep and ultrasteep spectra is higher
among the radio sources identified with galaxies than among those identified
with starlike objects.

Table~\ref{oi} lists the results of identification of 318 objects\footnote{
There are only two faint diffuse objects among the  320 radio sources, and
these come from the NVSS survey. These sources are absent on the FIRST survey
maps and hence cannot be identified.}
depending on the morphological type of the radio source.
Note that double sources (D, DC, DD) were identified mostly with galaxies.
The highest rate of failed identifications  is among the point sources (C).
The fraction of successful identifications for  CJ, CL, and DC-type objects
is higher and their optical counterparts are brighter than in case of point
and double radio sources, and there must be more nearby objects among the CJ,
CL, and DC-type sources. Distant objects must make up a significant fraction
of unidentified double  (D) and point (C) radio sources.

\subsection{Comparison of Two Flux-Density-Complete Samples}

In our first paper~\cite{Zhel} we compare the properties of radio sources in
two flux-density-complete samples of the RC catalog in the central part of
the  ``Cold'' survey. One of the samples includes 130 sources
located within the $\Delta H\le |5\arcmin|$ strip about the center of the
beam and with flux densities $S_{3.9\,GHz}\ge11$\,mJy.
The second sample contains 117 sources with $\Delta H\le |10\arcmin|$ and
$S_{3.9\,GHz}\ge30$~mJy.
These samples partially overlap in terms of objects. For the sake of brevity,
hereafter we refer to the first and second samples as  ``1S'' and ``2S'',
respectively.

\begin{table*}[tbp]
\setcaptionmargin{0mm} \onelinecaptionstrue
\captionstyle{flushleft}
\caption{Results of identification depending on the  $\alpha_{1.4\,-\,4.85\,GHz}$
spectral index of the radio source ($S\sim\nu^{-\alpha}$) for galaxies
(GALAXY), stellar objects (STAR), objects of unknown type (UNKNOWN), and
unidentified sources (Empty Field or EF)
}
\label{gsef}
\bigskip
\begin{tabular}{l|c|c|c|c|c}
\hline
~Type of identification~ & ~I ($\alpha<-0.1$)~ & ~F ($-0.1\ge\alpha<0.5$)~ & ~S ($0.5\ge\alpha<1$)~ & ~U ($\alpha\ge1$)~ & ~Data unavailable~ \\
	   & (29)              & (77)                    & (142)                & (31)             & (41)       \\
\hline
~GALAXY (157)       & 10 & 26 & 86 & 14 & 21 \\
~STAR   (88)        & 14 & 36 & 28 & 5  & 6  \\
~EF     (68)        & 5  & 12 & 25 & 11 & 14 \\
~UNKNOWN (6)        & -- & 2  & 3  & 1  & -- \\
\hline
\end{tabular}
\end{table*}

\begin{table*}[tbp]
\setcaptionmargin{0mm} \onelinecaptionstrue
\captionstyle{flushleft}
\caption{Results of the optical identification of the objects of the RC
catalog for radio sources of different morphological types. The counts do
not include four sources: two multicomponent objects, which we could not
attribute to any of the types mentioned below, and another two sources
absent in the FIRST survey.
We give the median values for LAS and~$m_{r}$
}
\label{oi}
\begin{center}
\begin{tabular}{ l | c | c | c | c | c | c | c | c | c | c }
\hline
~Type       & ~$N_{obj}~$ ~& $LAS$,~       & ~+~ & ~?~  & ~EF~ & ~STELLAR~ & ~$m^{stellar}_{r}$,~ & ~GALAXY~ & $m^{galaxy}_{r}$,~ & ~UNKNOWN~ \\
       &           & ~arcsec    &     &      &      &           &  mag                   &          &  mag                 &         \\
\hline
~C          & 123       & 1.4       & 78  & 10 & 35 & 39 & 20.23 & 45 & 21.04 & 4  \\
~CL         & 17        & 11.0      & 6   & -- &  1 & 6  & 19.64 & 9  & 16.00 & -- \\
~CJ         & 28        & 6.4       & 22  & 3  & 3  & 14 & 20.33 & 11 & 20.61 & -- \\
~T          & 19        & 34.9      & 17  & -- & 2  & 6  & 18.75 & 10 & 17.78 & 1  \\
~D, DC, DD~ & 129       & 17.5      & 93  & 11 & 25 & 22 & 19.75 & 81 & 21.30 & 1  \\
\hline
\end{tabular}
\end{center}
\end{table*}

\begin{table*}[tbp]
\setcaptionmargin{0mm} \onelinecaptionstrue
\captionstyle{flushleft}
\caption{Results of optical identification of two flux-density-complete
samples (1S and 2S) in the central part of the ``Cold'' survey
}
\label{1s2s}
\bigskip
\begin{tabular}{l|c|c|c|c|c|c}
\hline
~Sample~  & +           & ?           & EF          & ~GALAXY~    & STAR        & ~UNKNOWN~ \\
\hline
~1S (106)~ & ~90 (70\%)~ & ~16 (12\%)~ & ~24 (18\%)~ & 68 (64\%)   & ~36 (34\%)~ & 2 (2\%) \\
\hline
~2S (96)~  & ~84 (72\%)~ & ~12 (10\%)~ & ~21 (18\%)~ & 59 (61\%)   & ~35 (36\%)~ & 2 (2\%) \\
\hline
\end{tabular}
\end{table*}

\begin{table*}[tbp]
\setcaptionmargin{0mm} \onelinecaptionstrue
\captionstyle{flushleft}
\caption{Results of optical identification of two flux-density-complete samples (1S and 2S) for flat- and steep-spectrum
sources (in different intervals of  $\alpha_{1.4-4.85\;GHz}$ indices)
}
\label{spind}
\bigskip
\begin{tabular}{l|c|c|c|c|c|c|c}
\hline
~Sample/            & ~Fraction,~ & ~$S_{1.4\;GHz}^{med}$,~ & ~$S_{4.85\;GHz}^{med}$,~ & ~EF,~   & ~STELLAR,~ & ~GALAXY,~ & ~UNKNOWN,~ \\
~$N_{obj}$           & \%   & mJy                  & mJy                   & \%   & \%      & \%     & \%    \\
\hline
~1S ($\alpha$<0.5)   & 37   & 32.7  & 18  & 19   & 33  & 46  & 2 \\
~48                  &      &       &     &      &     &     &   \\
\hline
~1S ($\alpha\ge0.5$) & 63   & 82.3  & 30  & 18   & 25  & 56  & 1 \\
~82                  &      &       &     &      &     &     &   \\
\hline
~2S ($\alpha$<0.5)   & 28   & 52.5  & 46  & 12   & 52  & 33  & 3 \\
~33                  &      &       &     &      &     &     &   \\
\hline
~2S ($\alpha\ge0.5$) & 72   & 146.5 & 55  & 20   & 21  & 69  & 1 \\
~84                  &      &       &     &      &     &     &   \\
\hline
\end{tabular}
\end{table*}

\begin{table*}[tbp]
\setcaptionmargin{0mm} \onelinecaptionstrue
\captionstyle{flushleft}
\caption{Results of optical identification of two flux-density-complete
samples (1S and 2S) depending on the classification of objects
(G -- ``GALAXY''; S -- ``STAR'') and spectral index. The columns give the
median values
}
\label{spindo}
\bigskip
\begin{tabular}{l|l|c|c|c|c|c|c|c|c|c}
\hline
~Sample~          & ~Type,~  & ~$S_{4.85\,GHz}$,~ & ~LAS,~         & $\alpha$ & Size$_{opt}$, & $m_{u}$, & $m_{g}$, & $m_{r}$, & $m_{i}$, & $m_{z}$, \\
	   & \%   & mJy             & arcsec   &          & arcsec  & mag        & mag        &  mag       & mag        & mag         \\
\hline
~1S                & G (46) & 18 & 2.98 & ~0.19~ & 1.91 & ~22.42~ & ~21.12~ & ~20.04~ & ~19.48~ & ~19.14~ \\
~($\alpha<0.5$)~   & S (33) & 18 & 2.24 & ~0.24~ & 1.29 & ~21.03~ & ~20.69~ & ~20.28~ & ~19.82~ & ~19.83~ \\
\hline
~1S                & G (56) & 33 & 16.6 & ~0.80~ & 2.42 & ~22.66~ & ~22.45~ & ~21.17~ & ~20.23~ & ~19.31~ \\
~($\alpha\ge0.5$)~ & S (24) & 30 & 9.7  & ~0.67~ & 1.27 & ~20.56~ & ~20.03~ & ~19.51~ & ~19.41~ & ~19.39~ \\
\hline
~2S                & G (33) & 51 & 1.72 & ~0.25~ & 1.72 & ~22.85~ & ~19.72~ & ~18.49~ & ~18.10~ & ~17.93~ \\
~($\alpha<0.5$)~   & S (52) & 62 & 2.24 & ~0.13~ & 1.21 & ~21.14~ & ~20.38~ & ~19.98~ & ~19.75~ & ~19.39~ \\
\hline
~2S                & G (57) & 55 & 15.0 & ~0.84~ & 2.06 & ~22.73~ & ~22.77~ & ~21.51~ & ~20.34~ & ~20.04~ \\
~($\alpha\ge0.5$)~ & S (21) & 79 & 6.41 & ~0.78~ & 1.31 & ~22.16~ & ~20.24~ & ~20.13~ & ~19.36~ & ~19.20~ \\
\hline
\end{tabular}
\end{table*}

Table~\ref{1s2s} lists the comparative results of their optical
identification. The fraction of unidentified radio sources is the same in
both samples.

We subdivided each sample into two groups by their 1.4--4.85\,GHz spectral
indices. One group includes objects with flat spectra (\mbox{$\alpha<0.5$}), and the
other, sources with steep  (\mbox{$\alpha\ge0.5$}) spectra. We then compared the
results of identification  (see Tables~\ref{spind} and \ref{spindo}).
In both samples sources with flat spectra are more compact in terms of their
angular sizes both in the optical domain and at radio frequencies. The sources
with steep spectra are most often identified with galaxies both in the first
and second samples. The fraction of identifications with starlike objects is
higher for sources with flat spectra, and, moreover, the number of such
objects is almost twice higher than that of galaxies in the brighter
2S sample. Starlike objects with flat spectra form a separate group, because
they differ from other objects in both samples by their color indices.

Our two samples are, according to recent radio-source counts, dominated by
radio-loud AGNs. Thus most of the sources with flux densities
$S_{1.4\,GHz}\ge100$\,mJy are radio-loud AGNs with radio luminosities
above $2\times 10^{25}WHz^{-1}$ \cite{Willot}, i.e., FR~II-type sources
\cite{Fanaroff}, whereas fainter objects ($1<S_{1.4\,GHz}\le100$\,mJy) are
dominated by FR~I-type radio sources with luminosities below the given limit
\cite{Windhorst}. The subsample of identified radio sources maintains the
same galaxy-to-starlike object ratio (of about $2:1$),
despite the fact that the  1S sample is deeper in flux density terms than
the  2S sample. We assume that the overwhelming majority of starlike objects
are quasars as photometric and spectroscopic SDSS data suggest.

\subsection{Redshifts of Radio Sources}

Spectroscopic data are available in SDSS for 58 radio sources of our
identification list. These are 28 quasars and 30 galaxies. We found the radio
sources identified with galaxies to be rather nearby objects with
the median redshift and magnitude of \mbox{$Z_{galaxy}^{median}=0.20$} and
$m_{r}^{median}=18.7^{m}$, respectively, whereas those  identified with
quasars to be distant sources ($Z_{qsr}^{median}=1.76$). The latter are
brighter than galaxies by almost one magnitude ($m_{r}^{median}=17.1^{m}$).
These very objects are in the group of compact radio sources with flat
spectra, which we pointed out above. The most distant of the identified radio
sources are quasars with redshifts $Z>1.7$; they have flat or inverse radio
spectra. These are point radio sources (C) with angular sizes
$LAS\sim2\arcsec$, which are unresolvable in the FIRST survey.

We gathered the data on the redshifts and apparent magnitudes of the RC
catalog radio sources studied including the SS sample to analyze the
redshift--$m_{R}$ relation (see Fig. ~\ref{Zr}).
The plot is based on the data for 151 radio sources of the  RC catalog:
109 galaxies and 42 quasars.
These include the 58 mentioned above objects, 72 sources of the SS sample of
the RC catalog with the redshifts measured at the 6-m
telescope~\citep{Parijskij1,Afanasev,Soboleva2,Kopylov1,Kopylov2},
and 21 galaxies from among the objects identified as a result of this work.
For the latter we use the photometric redshifts from the SDSS
database~\cite{Csabai}, because the photometric redshifts in the
\mbox{$Z_{phot}<0.5-0.6$} interval for galaxies with \mbox{$m_{r}<20^{m}$}
and with redshift errors \mbox{$err_{Z_{phot}}<0.02-0.03$} agree well
with the empirical Z--R distribution for objects with known spectroscopic $Z$.

The redshift--magnitude relation\footnote{
We transform  $m_{r}$ into $m_{R_{c}}$ by the formula from
Jordi et al.~\cite{Jordi}.}
for radio galaxies shows up conspicuously out to $Z\sim1$ and persists at
greater redshifts, whereas quasars exhibit no such relation. The Z--R
diagram for radio galaxies at \mbox{$Z\ge1.5$} forms two branches,
which are possibly due to two groups of radio galaxies with different
luminosities.

\section{COMPARISON OF TWO-FREQUENCY SPECTRAL INDICES OF RADIO GALAXIES}

In the last two decades deep enough all-sky radio surveys have been conducted
at various frequencies. It goes without saying that the radio loud objects at
large redshifts detected in these surveys remain unidentified due to the lack
of efficient methods of selecting candidate counterparts for
deep objects. The use of multifrequency radio data may help to refine the
radio-source selection criteria aimed at detecting distant objects.

Our list of 143 sufficiently bright sources\footnote{
We give for these sources the median flux densities at the following
frequencies:
$S_{74\,MHz}$=1070\,mJy, $S_{365\,MHz}$=450\,mJy, $S_{1.4\,GHz}$=75\,mJy,
$S_{4.85\,GHz}$=41\,mJy.}
contains the data in the 74--4850\,MHz frequency interval given from the
VLSS, TXS, NVSS, and GB6 catalogs. We estimate the flux densities of some of
the objects by analyzing their images in the VLSS and GB6
surveys with the allowance for the coordinate dependent limiting sensitivity
of the maps.

We compare the two-frequency spectral indices of these sources and those of
the radio galaxies with $Z>3$ known from the literature. We took the
redshifts of 33 radio galaxies with $Z>3$ and the flux densities at 74 and
365\,MHz, 1.4 and 4.85\,GHz from~\cite{Kopylov1}, the list of powerful radio
galaxies given by Miley and De~Breuck~\cite{Milley}, and from the NED and
Vizier databases.

\begin{figure*}[tbp]
\setcaptionmargin{5mm}
\onelinecaptionsfalse
\includegraphics[width=12cm,bb=33 35 467 400,clip]{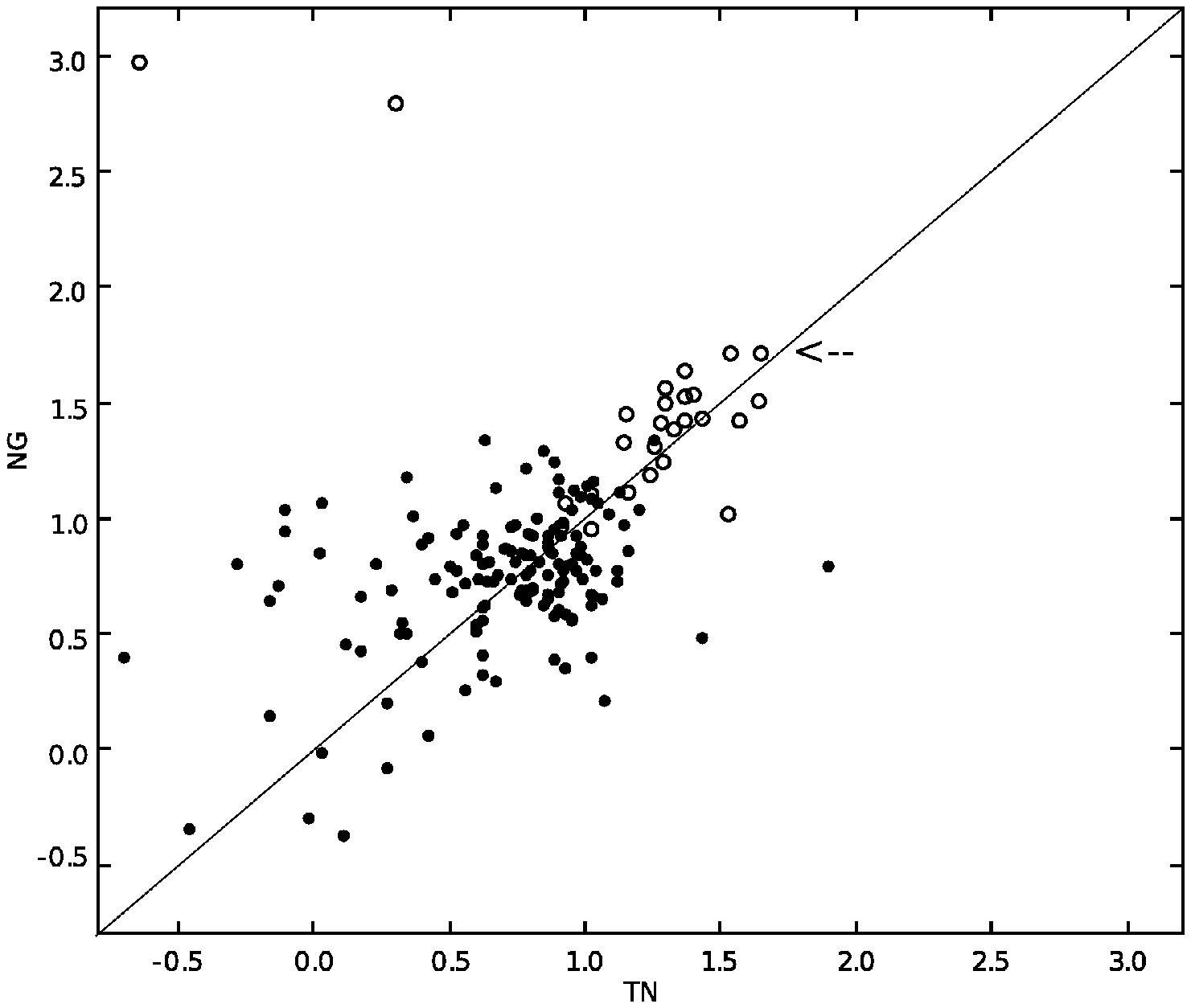}
\captionstyle{normal}
\caption{Comparison of two-frequency spectral indices
$\alpha_{365-1400\,MHz}$ (TN) and $\alpha_{1.4-4.85\,GHz}$ (NG) for
143 bright sources from the list studied (filled circles) and for known
distant radio galaxies with $Z>3$ (open circles). Also shown is the
equal-index line. The arrow indicates the most distant known radio galaxy
($Z=5.2$).
   }\label{TNNG}
\end{figure*}

We compare the  $\alpha_{74-365\,MHz}$ and $\alpha_{365-1400\,MHz}$;
$\alpha_{365-1400\,MHz}$ and $\alpha_{1.4-4.85\,GHz}$, as well as
$\alpha_{74-365\,MHz}$ and $\alpha_{1.4-4.85\,GHz}$ spectral indices  for the
RC-catalog radio sources and radio galaxies with redshifts $Z>3$ studied
in this paper. Figure~\ref{TNNG} shows by way of an example a comparison of
spectral indices. Two-frequency spectral indices for radio galaxies with
large redshifts differ from the corresponding indices for most of our sources.
Thus radio galaxies with $Z>3$ fall within the region determined
by the following constraints on spectral indices:
\begin{itemize}
\item
$\alpha_{74-365\,MHz}$ > 0.5;
\item
$\alpha_{365-1400\,MHz}$ > 0.9;
\item
$\alpha_{1.4-4.85\,GHz}$ > 0.7.
\end{itemize}

We selected from our list 18 radio sources with spectral indices fall within
the same area as the spectral indices of galaxies with large redshifts.
Thirteen of these sources appear in the SS sample of the RC catalog. Nine of
these 18 sources have their redshifts measured by the spectroscopic data
obtained with the 6-m telescope of the Special Astrophysical Observatory of
the Russian Academy of Sciences~\cite{Kopylov2}. Their median magnitudes,
angular sizes, and redshifts are equal to  $m_{R}=22.6^{m}$,
\mbox{$LAS=20.1\arcsec$,} and $Z$=0.82, respectively. No redshifts are known
for the remaining nine objects, which are fainter than
$m_{R}^{median}=23.4^{m}$ and have smaller angular sizes
($LAS^{median}=7.6\arcsec$): RC~J0820+0454, RC~J0945+0454, RC~J1347+0441,
RC~J1439+0455 (these sources are in the SS sample of the RC catalog)
and RC~J1251+0446, RC~J1357+0507, RC~1434+0445, RC~J1456+0456, and
RC~J1607+0438. The empirical Z--R relation for the RC catalog
(Fig. ~\ref{Zr}) allows the redshifts of these objects to be estimated as
$Z\ge1.5$. Their small angular sizes are consistent with this estimate,
however, only spectroscopic observations may finally confirm our hypotheses.

\section{CONCLUSIONS}

\begin{figure*}[tbp]
\setcaptionmargin{5mm}
\onelinecaptionsfalse
\includegraphics[width=15cm]{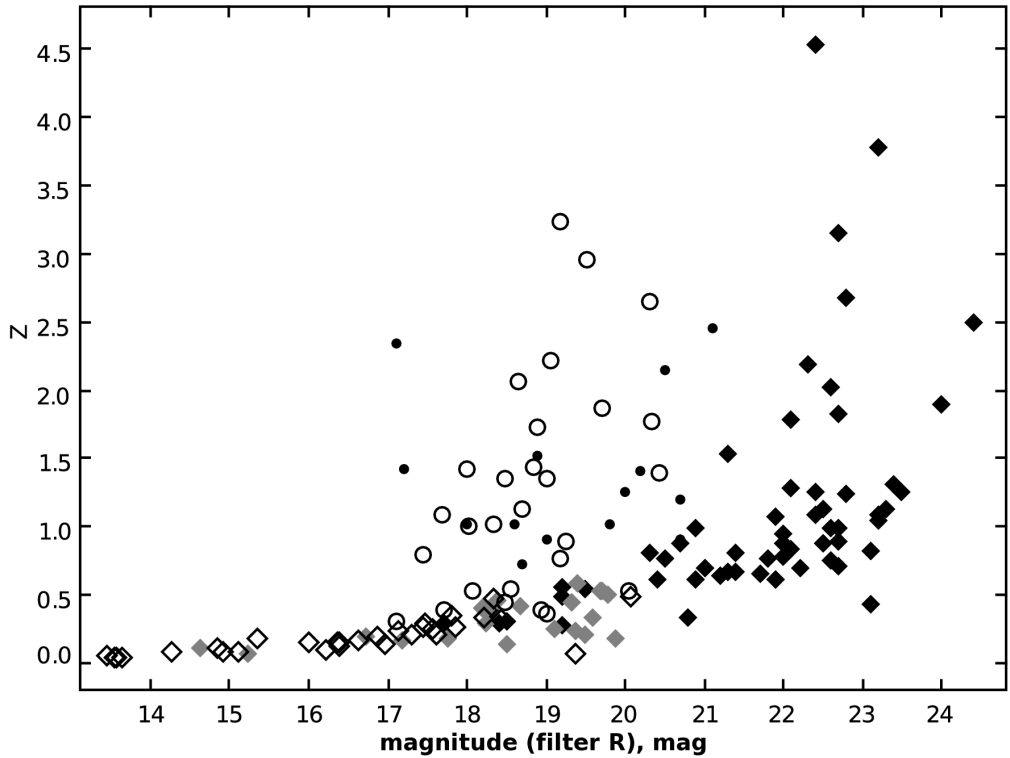}
\captionstyle{normal}
\caption{The  $m_{R}$ magnitude--redshift relation for the radio sources of
the RC catalog. The diamond signs and circles correspond to the galaxies and
quasars, respectively  (the filled diamond signs and circles indicate the
redshifts Z measured at the 6-m telescope, and the redshifts of the objects
shown by other symbols are adopted from the SDSS database).
   }\label{Zr}
\end{figure*}

The software for the analysis and visualization of observational data
developed in the past decade, and free access to modern surveys offer new,
hitherto unavailable opportunities for the study of celestial objects. A the
same time, the new tools impose new requirements both on the formulation of
the problems and on the methods of investigation, which in technological
solutions are increasingly dependent on information technologies and, first
and foremost, on database management systems and web services.

Mass identification of sources from various surveys and catalogs is of real
interest for the astronomers. Automating the identification process by using
already operational virtual observatory web services, which appears to be
rather easy to implement, is still far from perfect. This especially concerns
the radio data because of the wide range of angular resolution, limiting
sensitivity, coordinate accuracy of the catalogs, and the nature of radio
sources.

The coordinate accuracy of the NVSS and FIRST surveys allows them to be
automatically cross-identified with optical surveys. A number of authors
report the results of identification with the APM~\cite{McMahon} and SDSS
\cite{Ivezic} optical surveys.  The resulting rate of successful
identifications is rather low. It was equal to  24\% and 27\% in the former
and latter cases, respectively. If the coordinate accuracy of the catalog is
lower than that of the NVSS and SDSS surveys, as, e.g., in  GB6, the rate of
successful identifications is even lower---of about $0.2\%$ \cite{Obric}.
The number of FIRST sources identified with SDSS is about  30--40\%
of all objects in the catalog. These are single-component sources with the
offset between the optical and radio coordinates no greater than  $2\arcsec$.

Cross-identification services operate by the simple algorithm searching for
the nearest object to the center of the given region. This algorithm
efficiently identifies single-component radio sources, but handles poorly
sources with more complex structure. The efficiency of optical identification
of radio sources delivered by the existing cross-identification algorithms
does not exceed 30\%.

We decided to analyze the RC-catalog sample in order to understand to which
extent automatic identification of radio sources can be performed in case
where the angular resolution and coordinate accuracy of the catalog are
insufficient for optical identification, what fraction of sources can be
identified, and to find out the problems and restrictions of such a procedure.

Due to the insufficient coordinate accuracy of the RC catalog the procedure
of identification of the sample in the area overlapping with the  SDSS and
FIRST surveys consists of two stages. We first refined the coordinates of
the radio sources of the catalog by cross-identifying them with the sources
in other radio catalogs with better coordinate accuracy (mostly NVSS, which
we used as our reference catalog). In the doubtful cases we used not  only
the data from the catalogs, but also the images provided in the radio surveys.
We performed optical identification using refined radio coordinates and
contour radio maps based on the FIRST survey images. We refined the
classification of sources for it to include seven types, because we concluded
that the classification including FRI, FRII, and point sources failed to
fully reflect the structural variety of the sources of the FIRST survey.

We identified 320 radio sources of the  RC catalog with sources from other
radio catalogs and with optical objects. We measured the angular sizes of
these objects and counted the number of their components listed in the FIRST
catalog.
We found the ratio of the radio sources with \mbox{one-, two, three-,} and four or
more components to be of about 10:5:2:1.

It follows from these counts that simple cross-identification algorithms
(based on the nearest neighbor search) are the most efficient ones for
single-component sources ($56\%$ of the list) provided the optical survey is
deep enough.
The algorithms must be modified if they are going to be used for identifying
double sources, which make up for about one third of the list. One fifth of
the list can be identified by inspecting optical and radio images or by
applying cross-identification algorithms involving elements of pattern
recognition. We identified our sample with SDSS and found optical candidates
for about  $75\%$ of single-component sources, which make up for $33\%$ of
the entire list.

Cross-identification depends essentially on the adopted search radius, which
is a parameter characterizing the two catalogs compared. Thus the optimum
search radius for FIRST and SDSS catalogs for single-component sources is
$2\arcsec$. Identification is unlikely if the the separation between the
radio source and optical object is greater than this radius.

We compared the threshold values for the maxi\-mum-likelihood ratio functions
for three groups of sources in our sample---point, double, and non-point
sources with the coincident positions of the optical object and
radio-emission peak, to obtain our own estimate for the search region for
optical objects in the FIRST survey. The search radius for point sources is
of about $1.9\arcsec$, which almost coincides with the search radius for
FIRST and SDSS mentioned above. The search radius for non-point and double
sources is of about $3.6\arcsec$ if the search region center is set to
coincide with the center of the radio source as we define it in our first
paper~\cite{Zhel}.

We found candidate optical identifications for almost  $80\%$ of the radio
sources of the list studied and no candidates for the remaining  $20\%$ of
the sources, i.e., optical objects proved to be fainter than the limiting
magnitude of the SDSS survey ($r=22.6^{m}$). The galaxy-to-starlike object
ratio among identified sources is of about  $2:1$, and starlike objects are
most probably quasars judging by the photometric and spectroscopic SDSS data.

The distribution of the results of identifications by morphological type of
the radio source is such that most of the optical candidates have been found
for CJ, CL, and T-type sources; smaller amount of optical candidates could be
found for double radio sources, and even less for point sources. Double
sources are mostly identified with galaxies. CL and T-type sources are
identified with brighter optical objects than  CJ-type, double,
and point radio sources. A large fraction of radio sources identified with
galaxies have steep and ultrasteep spectra in the \mbox{1.4--4.85\,GHz} frequency
interval compared to starlike objects, however, the fraction of objects
with steep and ultrasteep spectra is even higher among the radio sources
in ``empty fields''. Some of the unidentified point sources and objects with
ultrasteep spectra in the 1.4--4.85\,GHz frequency interval may have
large~$Z$.

The structure of a radio source may bear information about recurrent phases
of its activity in the host galaxy. Such systems include ``winged'' or
``X-shaped'' radio sources~\cite{Liu,Cheung},
``double-double'' radio galaxies~\cite{Lara}, and triple sources (T) with two
components straddling relatively bright, unresolved
cores~\cite{Marecki}. According to FIRST survey maps, about \mbox{$12-17\%$}
of the radio sources of the sample studied can be classified as ``X-shaped'',
``double-double'', or triple, i.e., as sources with recurrent activity phases.

We analyzed the physical parameters of the identified radio sources by
comparing two flux-density-complete samples from the central part of the
survey \cite{Zhel}: one with \mbox{$S^{lim}_{3.9\,MHz}\ge10$\,mJy} and another with
\mbox{$S^{lim}_{3.9\,MHz}\ge30$\,mJy.} We found the fraction of
unidentified radio sources to be the same in both samples,
and the galaxy-to-starlike object ratio to be the same (of about 2:1) in
both subsamples of identified sources.
However, the two samples differ in the fraction of sources with flat and
steep spectra. The deeper sample contains more objects with flat and inverse
spectra and less objects with steep and ultrasteep spectra.

In both samples sources with flat spectra are more compact in terms of
angular size both on optical and radio images. Steep-spectrum sources are
more often identified with galaxies, and flat-spectrum sources---with
starlike objects. Note that in the  2S sample the number of identifications
of flat-spectrum sources with starlike objects is almost twice the number of
identifications with galaxies. In both samples starlike objects with flat
spectra differ in color from the galaxies with flat spectra, whereas no such
differences are observed for sources with steep spectra.

Optical identification with SDSS revealed a group of nearby radio galaxies
with $Z_{sp}^{median}$=0.20 and $m_{r}^{median}=18.7^{m}$ and a group of
rather distant radio-loud quasars with $Z_{sp}^{median}$=1.76,
$m_{r}^{median}=17.1^{m}$. Note that quasars with $Z>1.7$ are point sources
with angular sizes $LAS^{median}\sim2\arcsec$.
Quasars have flat and inverse radio spectra and, possibly,  make up the
above-mentioned group of radio sources with flat spectra and with a spectral
distribution that is different from that of other sources.

The steepness of the spectrum of a radio source from low to high frequencies
is widely used as a criterion for selecting possible distant objects. We
compared the spectral indices for the sources in our list, for those with
the data at 74, 365\,MHz and 1.4, 4.85\,GHz are available, with 33 known
radio galaxies with $Z>3$. We found that distant objects occupy a
certain region of index values on two-frequency indices plots. We selected
from the list studied the radio sources with spectral indices fall
within the same region as the spectral indices of galaxies at large redshifts.
Two thirds of these sources belong to the SS sample of the RC catalog and
most of them have their redshifts measured
($Z^{median}$=0.82)~\cite{Kopylov2}, whereas objects with unknown $Z$ are
fainter and have smaller angular sizes. The last group in terms of apparent
magnitude occupies the domain of RC catalog sources with $Z\ge1.5$
on the Z--R diagram  (Fig.~\ref{Zr}), i.e., these objects are most likely
rather distant galaxies. A comparison of the spectral indices of the sources
studied with those of distant galaxies led us to conclude that the shape of
the radio spectrum in the  74--4850\,MHz interval provides additional
information that may help to refine the choice of candidate distant objects.

The tables with the results of the optical cross-identification of 320 radio
sources of the RC catalog with SDSS, \mbox{USNO-B1}, and 2MASS are available in
electronic form along with their description from
{\tt http://www.sao.ru/hq/zhe/RCoiRes.html}.

\begin{acknowledgements}
The large amount and inhomogeneous nature of the data, which includes eight
catalogs and four surveys, would be impossible to prepare and analyze without
new software tools developed in accordance with IVOA standards, namely
Aladin~\cite{Bonnarel}, Vizier~\cite{Ochsenbein}, TOPCAT~\cite{Taylor}, and
CasJobs~\cite{Thakar}.

This work was supported by the Russian Foundation for Basic Research
(grant No 06-07-08062).
\end{acknowledgements}

\end{document}